\documentclass[letterpaper]{article} 
\usepackage{aaai2026}  
\usepackage{times}  
\usepackage{helvet}  
\usepackage{courier}  
\usepackage[hyphens]{url}  
\usepackage{graphicx} 
\usepackage{natbib}  
\usepackage{caption} 
\usepackage{algorithm}
\usepackage{algorithmic}

\usepackage{newfloat}
\usepackage{listings}
\DeclareCaptionStyle{ruled}{labelfont=normalfont,labelsep=colon,strut=off} 
\floatstyle{ruled}
\newfloat{listing}{tb}{lst}{}
\floatname{listing}{Listing}
\usepackage{amssymb}
\usepackage{bm}
\usepackage{makecell}
\usepackage{threeparttable}
\usepackage{amsmath}
\usepackage{booktabs}
\usepackage{bbding}
\usepackage{subfigure}
\usepackage{hyperref}

\makeatletter
\def\thanks#1{\protected@xdef\@thanks{\@thanks
        \protect\footnotetext{#1}}}
\makeatother

\nocopyright 

\title{Rethinking Flow and Diffusion Bridge Models for Speech Enhancement\thanks{Accepted by the 40th AAAI Conference on Artificial Intelligence (AAAI-26).}}
\author{
    Dahan Wang\textsuperscript{\rm 1,2}, Jun Gao\textsuperscript{\rm 1,2}, Tong Lei\textsuperscript{\rm 3}, Yuxiang Hu\textsuperscript{\rm 2}, Changbao Zhu\textsuperscript{\rm 2}, Kai Chen\textsuperscript{\rm 1,2}, and Jing Lu\textsuperscript{\rm *1,2} 
}
\affiliations{
    \textsuperscript{\rm 1}Key Laboratory of Modern Acoustics, Institute of Acoustics, Nanjing University, Nanjing, China\\
    \textsuperscript{\rm 2}NJU-Horizon Intelligent Audio Lab, Horizon Robotics, Beijing, China\\
    \textsuperscript{\rm 3}Tencent AI Lab, Shenzhen, China\\
    \textsuperscript{\rm *}Corresponding author: lujing@nju.edu.cn
}

\usepackage{bibentry}

\begin{document}

\maketitle

\begin{abstract}
Flow matching and diffusion bridge models have emerged as leading paradigms in generative speech enhancement, modeling stochastic processes between paired noisy and clean speech signals based on principles such as flow matching, score matching, and Schr\"odinger bridge. In this paper, we present a framework that unifies existing flow and diffusion bridge models by interpreting them as constructions of Gaussian probability paths with varying means and variances between paired data. Furthermore, we investigate the underlying consistency between the training/inference procedures of these generative models and conventional predictive models. Our analysis reveals that each sampling step of a well-trained flow or diffusion bridge model optimized with a data prediction loss is theoretically analogous to executing predictive speech enhancement. Motivated by this insight, we introduce an enhanced bridge model that integrates an effective probability path design with key elements from predictive paradigms, including improved network architecture, tailored loss functions, and optimized training strategies. Experiments on denoising and dereverberation tasks demonstrate that the proposed method outperforms existing flow and diffusion baselines with fewer parameters and reduced computational complexity. The results also highlight that the inherently predictive nature of this generative framework imposes limitations on its achievable upper-bound performance. 
\end{abstract}

\begin{links}
    \link{Appendix, code, and audio samples}{https://github.com/Dahan-Wang/Rethinking-Flow-and-Diffusion-Bridge-Models-for-Speech-Enhancement}
\end{links}

\section{Introduction}

Deep learning-based methods have achieved remarkable success in speech enhancement (SE), which aims to recover clean speech from noisy observations. These methods can be broadly categorized into predictive (discriminative) and generative frameworks. Predictive models \cite{yin2020phasen, zheng2021interactive} learn a direct mapping from noisy signals to clean speech, whereas generative methods model the distribution of clean speech conditioned on its noisy counterpart. Recently, various generative paradigms have been extensively explored, including generative adversarial networks (GANs) \cite{fu2019metricgan}, variational autoencoders (VAEs) \cite{fang2021variational}, self-supervised learning (SSL) models \cite{wang2024selm}, and diffusion models \cite{tai2023dose, lei2024shallow, richter2024diffusion, liu2024latent, li2025complex}. These generative approaches consistently demonstrate promising performance and robust generalization across diverse unseen acoustic scenarios.

In flow and diffusion-based models, SE is naturally formulated as a conditional generation task \cite{tai2023revisiting}. One of the most widely adopted paradigms is to introduce noisy speech into both the prior distribution and the conditional probability path. Task-adapted score-based diffusion models \cite{lemercier2025diffusion} achieve this by designing the drift term of continuous-time stochastic differential equations (SDEs) based on either the Ornstein-Uhlenbeck (OU) process \cite{richter2023speech} or the Brownian bridge (BB) \cite{lay2023reducing}, resulting in diffusion processes with means interpolating between clean and corrupted signals. More recently, the tractable Schr\"odinger bridge (SB) framework \cite{chen2023schrodinger}, which is also referred to as the denoising diffusion bridge model (DDBM) \cite{he2024consistency}, has been proposed to build stochastic processes between Dirac noisy and clean data endpoints by optimizing path measures under boundary constraints. The SB model also incorporates a data prediction training strategy, achieving state-of-the-art (SOTA) performance compared to conventional diffusion models \cite{jukic2024schrodinger}. Additionally, the flow matching (FM) method has been extended to incorporate probability paths conditioned on noisy speech, enabling efficient sampling while maintaining strong SE performance \cite{korostik2025modifying, lee2025flowse}. These works have become the foundational basis for numerous advances in generative SE \cite{lemercier2023storm, lay2024single, richter2025investigating}.

The aforementioned methods  are grounded in distinct theoretical foundations, including score-based diffusion, Schr\"odinger bridge, and flow matching, which have yet to be unified under a common framework in the SE field. Additionally, the use of data prediction objectives in diffusion bridge models \cite{chen2023schrodinger} suggests their resemblance to predictive methods, which similarly estimate clean speech by implicitly learning distributional mappings between paired data. This connection, however, remains underexplored in prior work.

In this paper, we present a unified framework for flow and diffusion bridge models in SE, interpreting them as constructing different Gaussian probability paths between paired noisy and clean data. Then the sampling ordinary differential equations (ODEs) are derived through conditional flow matching, and extended to SDEs for both forward and backward processes. Notably, we show that all such models can be trained using a data prediction strategy. The fundamental difference among them lies in the design of mean and variance trajectories. Our analysis further reveals that each sampling step in a well-trained flow matching or diffusion bridge model is theoretically equivalent to predictive SE, and the final output is a weighted sum of these step-wise predictions. This suggests that these models, while generative in form, fundamentally operate as predictive models---explaining their effectiveness in single-step sampling and highlighting opportunity for improvement via predictive techniques.

Motivated by the above insights, we propose an enhanced bridge model that combines an effective probability path design with key strengths of the predictive paradigm. Specifically, we adopt a high-performance backbone \cite{wang2023tf}, and introduce a time embedding mechanism to effectively leverage the information encoded in the diffusion time. Moreover, we refine the data prediction loss to optimize the model training, and integrate a fine-tuning strategy \cite{lay2024single} for further performance gain. Experimental results reveal that the proposed model outperforms SOTA flow matching- and diffusion-based baselines while incurring markedly fewer parameters and reduced computational overhead. Furthermore, our findings highlight an upper-bound performance constraint imposed by the predictive nature of such generative frameworks.

Our main contributions are summarized as follows:
\begin{itemize}
\item \textbf{Unified Generative Framework:} We present a unified theoretical framework that encompasses existing flow and diffusion bridge models between paired data, including score-based diffusion, Schr\"odinger bridge, and flow matching, which are widely used generative approaches in SE.

\item \textbf{Predictive Equivalence Insight:}  We investigate the inherent equivalence between flow matching/diffusion bridge models and predictive methods, showing that these generative models share key mechanisms with predictive models. This insight provides practical guidance for model improvement and suggests that the predictive nature of such generative models may impose a ceiling on their performance. 

\item \textbf{Enhanced Bridge Model:} Our proposed enhanced bridge model incorporates advanced predictive strategies. Our model achieves significantly better performance and efficiency compared to existing flow and diffusion baselines.
\end{itemize}

\section{Related Work}

\subsection{Score-based Diffusion Models}

Score-based generative models \cite{welker2022speech, richter2023speech} describe the forward diffusion process through the forward SDE:
\begin{equation}
    \mathrm{d}\mathbf{x}_t = \mathbf{f}_t(\mathbf{x}_t,\mathbf{y})\mathrm{d}t+g_t\mathrm{d}\mathbf{w}_t,
    \label{eq.1}
\end{equation}
where $t \in [0, 1]$ denotes a continuous time variable, $\mathbf{x}_t \in \mathbb{C}^{F \times L}$ represents the state of the process, i.e. the reshaped spectral coefficient vector with $F$ frequency bins and $L$ frames, $\mathbf{y}$ is the noisy speech vector, $\mathbf{w}_t$ is a standard Wiener process, $\mathbf{f}_t(\cdot, \cdot)$ is the drift term, and $g_t$ is a scalar-valued diffusion coefficient. The initial condition of $\mathbf{x}_t$ is the clean speech $\mathbf{s}$. For the OU process, the drift term is defined as $\mathbf{f}_t(\mathbf{x}_t, \mathbf{y}) = \gamma (\mathbf{y} - \mathbf{x}_t)$, where $\gamma$ is the stiffness coefficient. For the BB process, the drift term is $\mathbf{f}_t(\mathbf{x}_t, \mathbf{y}) = \frac{\mathbf{y} - \mathbf{x}_t}{1 - t}$. For the diffusion coefficient, the variance-exploding (VE) schedule is commonly adopted, i.e., $g_t = \sqrt{c}k^t$. The combinations of these drift and diffusion terms are referred to as Ornstein-Uhlenbeck with variance exploding (OUVE) \cite{richter2023speech} and Brownian bridge with exponential diffusion coefficient (BBED) \cite{lay2023reducing}, respectively. The reverse SDE and its corresponding probability flow ODE (PFODE) are respectively given by
\begin{equation}
    \mathrm{d}\mathbf{x}_t = \left[\mathbf{f}_t(\mathbf{x}_t, \mathbf{y}) \! - \! g_t^2 \nabla_{\mathbf{x}_t} \log p_t(\mathbf{x}_t | \mathbf{s}, \mathbf{y}) \right] \mathrm{d}t + g_t \mathrm{d}\bar{\mathbf{w}}_t,
    \label{eq.2}
\end{equation}
\begin{equation}
    \mathrm{d}\mathbf{x}_t = \left[\mathbf{f}_t(\mathbf{x}_t, \mathbf{y}) - \frac{1}{2} g_t^2 \nabla_{\mathbf{x}_t} \log p_t(\mathbf{x}_t | \mathbf{s}, \mathbf{y}) \right] \mathrm{d}t,
    \label{eq.3}
\end{equation}
where $\mathrm{d}t$ represents a negative infinitesimal time step, $\bar{\mathbf{w}}_t$ is the reverse-time Wiener process, $p_t(\mathbf{x}_t | \mathbf{s}, \mathbf{y})$ denotes the conditional probability path (or perturbation kernel), and $\nabla_{\mathbf{x}_t} \log p_t(\mathbf{x}_t | \mathbf{s}, \mathbf{y})$ is the corresponding score function. The probability path has a Gaussian form defined by 
\begin{equation}
    p_t(\mathbf{x}_t | \mathbf{s}, \mathbf{y}) = \mathcal{N} \left( \mathbf{x}_t; \boldsymbol{\mu}_t(\mathbf{s}, \mathbf{y}) , \sigma_t^2\mathbf{I} \right),
    \label{eq.4}
\end{equation}
with its mean and variance determined by $\mathbf{f}_t$ and $g_t$. The score function can be obtained via denoising score matching:
\begin{equation}
    \nabla_{\mathbf{x}_t} \log p_t(\mathbf{x}_t | \mathbf{s}, \mathbf{y}) = - \frac{\mathbf{x}_t - \boldsymbol{\mu}_t}{\sigma_t^2},
    \label{eq.5}
\end{equation}
which is the training objective of the backbone network.

\subsection{Schr\"odinger Bridge}

The SB problem originates from the optimization of path measures with constrained boundaries. For dual Dirac distribution boundaries centered on paired clean and noisy speech, the SB solution can be expressed as a couple of forward-backward SDEs \cite{chen2023schrodinger}:
\begin{equation}
    \mathrm{d}\mathbf{x}_t = \left[ f_t\mathbf{x}_t - g^2_t\frac{\mathbf{x}_t - \bar{\alpha}_t\mathbf{y}}{\alpha_t^2\bar{\rho}_t^2} \right]\mathrm{d}t + g_t\mathrm{d}\mathbf{w}_t,
    \label{eq.6}
\end{equation}
\begin{equation}
    \mathrm{d}\mathbf{x}_t = \left[ f_t\mathbf{x}_t + g^2_t\frac{\mathbf{x}_t - \alpha_t\mathbf{s}}{\alpha_t^2\rho_t^2} \right]\mathrm{d}t + g_t\mathrm{d}\bar{\mathbf{w}}_t,
    \label{eq.7}
\end{equation}
with the corresponding probability path defined as
\begin{equation}
    p_t(\mathbf{x}_t | \mathbf{s}, \mathbf{y}) = \mathcal{N}\left( \frac{\alpha_t\bar{\rho}_t^2\mathbf{s} + \bar{\alpha}_t\rho_t^2\mathbf{y}}{\rho_1^2}, \frac{\alpha_t^2\bar{\rho}_t^2\rho_t^2}{\rho_1^2}\mathbf{I} \right),
    \label{eq.8}
\end{equation}
and the PFODE formulated as
\begin{equation}
    \mathrm{d}\mathbf{x}_t \mkern-4mu = \mkern-4mu \left[ f_t\mathbf{x}_t \mkern-4mu - \mkern-4mu \frac{1}{2} g^2_t\frac{\mathbf{x}_t - \bar{\alpha}_t\mathbf{y}}{\alpha_t^2\bar{\rho}_t^2} \mkern-4mu + \mkern-4mu \frac{1}{2} g^2_t\frac{\mathbf{x}_t - \alpha_t\mathbf{s}}{\alpha_t^2\rho_t^2} \right] \mathrm{d}t,
    \label{eq.9}
\end{equation}
where $f_t$ is the drift coefficient, $\alpha_t = \exp\left( 
\int_0^t f_\tau\mathrm{d}\tau \right)$, $\rho_t^2 = 
\int_0^t g^2_\tau \alpha_\tau^{-2} \mathrm{d}\tau $, $\bar{\alpha}_t = \alpha_t\alpha_1^{-1}$, and $\bar{\rho}_t^2 = \rho_1^2 - \rho_t^2$. This set of formulations can serve as a unified framework for all DDBMs between paired data \cite{he2024consistency}. For SE, a data prediction training strategy is widely adopted due to its performance advantages over score matching, that is, the network directly predicts the clean speech $\mathbf{s}$. Moreover, VE is the most commonly used schedule in SE, defined by setting $f_t = 0$ and $g_t = \sqrt{c}k^t$, which is referred to as SBVE \cite{jukic2024schrodinger}. In this paper, the SB model and the score-based diffusion models introduced in the previous subsection are collectively referred to as diffusion bridge models.

\subsection{Flow Matching}

A flow matching method for SE is defined by an ODE:
\begin{equation}
    \mathrm{d}\mathbf{x}_t = \mathbf{u}_t ( \mathbf{x}_t | \mathbf{s}, \mathbf{y} ) \mathrm{d}t,
    \label{eq.10}
\end{equation}
where $\mathbf{u}_t ( \mathbf{x}_t | \mathbf{s}, \mathbf{y} )$ denotes the conditional vector field. Unlike diffusion models, the sampling process in flow models proceeds forward in time, with $t = 1$ corresponding to the target data distribution. Restricting $\mathbf{x}_t$ to follow a Gaussian probability path, the conditional vector can be derived as
\begin{equation}
    \mathbf{u}_t(\mathbf{x}_t | \mathbf{s}, \mathbf{y}) = \frac{\sigma_t'}{\sigma_t}(\mathbf{x}_t-\boldsymbol{\mu}_t) + \boldsymbol{\mu}_t'.
    \label{eq.11}
\end{equation}
For SE tasks with paired clean and noisy data, following the optimal transport conditional FM (OT-CFM), the mean and variance of the probability path are set to $\boldsymbol{\mu}_t(\mathbf{s}, \mathbf{y}) = (1 - t) \mathbf{y} + t \mathbf{s}$ and $\sigma_t = (1 - t) \sigma_{\max} + t \sigma_{\min}$, respectively \cite{korostik2025modifying, lee2025flowse}.

\section{Methodology}

\subsection{A Unified Framework for Flow and Diffusion Bridge Models}

\subsubsection{Framework}
We define the probability path in Gaussian form, as given in Eq.~{(\ref{eq.4})}, with the mean specified as
\begin{equation}
    \boldsymbol{\mu}_t(\mathbf{x}_t | \mathbf{s}, \mathbf{y}) = a_t \mathbf{s} + b_t \mathbf{y}.
    \label{eq.12}
\end{equation}
which interpolates between the clean and noisy speech. Based on Eq.~{(\ref{eq.11})}, the corresponding ODE is derived as
\begin{equation}
    \frac{\mathrm{d}\mathbf{x}_t}{\mathrm{d}t} = \frac{\sigma_t'}{\sigma_t} \mathbf{x}_t + \left( a_t' - a_t \frac{\sigma_t'}{\sigma_t} \right) \mathbf{s} + \left( b_t' - b_t \frac{\sigma_t'}{\sigma_t} \right) \mathbf{y}.
    \label{eq.13}
\end{equation}
Following the SDE extension trick based on the Fokker-Planck equation \cite{holderrieth2025introduction}, the associated forward-backward SDEs are formulated as
\begin{equation}
    \mathrm{d}\mathbf{x}_t \! = \! \left[ \kappa_t^{\scriptscriptstyle +} \mathbf{x}_t \mkern-4mu + \mkern-4mu \left( a_t' \mkern-4mu - \mkern-4mu a_t \kappa_t^{\scriptscriptstyle +} \right) \mathbf{s} \mkern-4mu + \mkern-4mu \left( b_t' \mkern-4mu - \mkern-4mu b_t \kappa_t^{\scriptscriptstyle +} \right) \mathbf{y} \right] \! \mathrm{d}t \! + \! g_t\mathrm{d}\mathbf{w}_t,
    \label{eq.14}
\end{equation}
\begin{equation}
    \mathrm{d}\mathbf{x}_t \! = \! \left[ \kappa_t^{\scriptscriptstyle -} \mathbf{x}_t \mkern-4mu + \mkern-4mu \left( a_t' \mkern-4mu - \mkern-4mu a_t \kappa_t^{\scriptscriptstyle -} \right) \mathbf{s} \mkern-4mu + \mkern-4mu \left( b_t' \mkern-4mu - \mkern-4mu b_t \kappa_t^{\scriptscriptstyle -} \right) \mathbf{y} \right] \! \mathrm{d}t \! + \! g_t\mathrm{d}\bar{\mathbf{w}}_t,
    \label{eq.15}
\end{equation}
where
\begin{equation}
    \kappa_t^{\scriptscriptstyle \pm} = \frac{\sigma_t'}{\sigma_t} \mp \frac{g_t^2}{2\sigma_t^2}.
    \label{eq.16}
\end{equation}
The detailed derivation is provided in Appendix~\ref{A1}.

Based on the above framework, we interpret the core design principle of flow and diffusion bridge models as the construction of conditional probability paths between paired data, specifically through the design of $a_t$, $b_t$, and $\sigma_t$. Once the probability path is specified, the corresponding sampling equations can be directly obtained via Eqs.~{(\ref{eq.13})-(\ref{eq.15})}. This set of unified formulations enables a consistent description of various SE generative models without the need to start from the design of forward SDEs, as in score-based diffusion models, or to solve Kullback-Leibler-divergence optimization and partial differential equations, as required by the SB method. The parameters defining the probability paths in representative models are summarized in Table~\ref{Table1}. Detailed proofs of how these models are derived from our framework are provided in Appendix~\ref{A2}.

\begin{table}[t]
\renewcommand{\arraystretch}{1.2}
\centering
\small
\begin{threeparttable}
\begin{tabular}{l|ccc}
\Xhline{1pt}
Method & $a_t$ & $b_t$ & $\sigma_t$\\ 
\hline\hline
OUVE & $e^{-\gamma t}$ & $1 - e^{-\gamma t}$ & $\frac{c \left( k^{2t} - \mathrm{e}^{-2\gamma t} \right)}{2(\gamma + \log k)}$ \\
BBED & $1-t$ & $t$ & $c(1-t)E_t \tnote{*}$ \\
SB & $\alpha_t\bar{\rho}_t^2 / \rho_1^2$ & $\bar{\alpha}_t\rho_t^2 / \rho_1^2$ & $\alpha_t^2\bar{\rho}_t^2\rho_t^2 / \rho_1^2$ \\
OT-CFM & $t$ & $1-t$ & $(1 - t) \sigma_{\max} + t \sigma_{\min}$ \\
\Xhline{1pt}
\addlinespace[2pt]
\end{tabular}
\begin{tablenotes}
\item[*] $E_t = (k^{2t} - 1 + t) + \log(k^{2k^2}) \{ \text{Ei} \left[ 2(t-1) \log k \right] - \text{Ei} \left[ -2 \log k \right] \} (1-t)$, where $\text{Ei}[\cdot]$ denotes the exponential integral function \cite{bender2013advanced}.
\end{tablenotes}
\end{threeparttable}
\caption{\label{Table1} Probability path parameters of representative flow and diffusion bridge models for SE.}
\end{table}

\subsubsection{Diffusion Coefficient and Sampling Direction} Note that there are two important issues regarding the forward-backward SDEs that require further clarification. First, to derive the SDEs, the form of the diffusion coefficient $g_t$ must be specified. Theoretically, $g_t$ can be arbitrary, meaning that a single probability path may correspond to a family of SDEs with different diffusion coefficients. This is because, according to the Fokker-Planck equation, the effects of $g_t$ on the drift and diffusion terms cancel out, preserving the same underlying probability path \cite{holderrieth2025introduction}. In previous diffusion bridge models, the designed SDEs represent a specific, tractable case within this broader family with arbitrary $g_t$. The $g_t$ defined in these models is related to $\sigma_t$, and this relationship can be used to simplify the form of the resulting ODE and SDEs.

Second, our framework does not impose a fixed temporal direction for sampling. Instead, the direction is determined by the definitions of the path parameters. Typically, the sampling process starts at a point with mean $\mathbf{y}$ and ends at a point with mean $\mathbf{s}$ and zero variance. However, the assignment of these conditions to $t = 0$ or $t = 1$ is not fixed in advance, which is governed by the definitions of $a_t$, $b_t$, and $\sigma_t$. For diffusion bridge models, the sampling process proceeds in reverse time, meaning that the backward SDE (Eq.~(\ref{eq.14})) is used for sampling. In contrast, for flow matching models, the sampling proceeds in forward time; even when extended to the SDE form, the forward SDE is used for sampling.

\subsubsection{Training and Sampling}

According to Eqs.~{(\ref{eq.13})-(\ref{eq.15})}, the only unknown term during the sampling process is the clean speech $\mathbf{s}$. Therefore, the network can be trained using a data prediction strategy, where the clean speech $\mathbf{s}$ serves as the training target. $\mathbf{s}$ in these equations is replaced by the network's output during sampling. This strategy is particularly advantageous for SE tasks, as it allows the incorporation of auxiliary losses tailored to the characteristics of speech signals \cite{chen2023schrodinger, richter2025investigating}. Moreover, our framework enables the application of data prediction training to OUVE and BBED, which originally rely on score matching for optimization.

We recommend using a discretization method based on exponential integrators for sampling, as it introduces minimal discretization error \cite{chen2023schrodinger, he2024consistency}. For simplicity, we rewrite the ODE presented in Eq.~{(\ref{eq.13})} as $\frac{\mathrm{d}\mathbf{x}_t}{\mathrm{d}t} = \frac{\sigma_t'}{\sigma_t} \mathbf{x}_t + m_t \mathbf{s} + n_t \mathbf{y}$, which enables the corresponding discretized sampling equation to be expressed as
\begin{equation}
    \mathbf{x}_t = \frac{\sigma_t}{\sigma_r} \mathbf{x}_r + \sigma_t \left( \int_r^t \frac{m_\tau}{\sigma_\tau} \mathrm{d}\tau \right) \mathbf{s} + \sigma_t \left( \int_r^t \frac{n_\tau}{\sigma_\tau} \mathrm{d}\tau \right) \mathbf{y}.
    \label{eq.17}
\end{equation}
However, for certain models with complex parameterizations (such as OUVE and BBED), the integral in this expression may not yield a tractable closed-form solution, making the exponential integrator method difficult to apply to these methods. The detailed derivation and further discussion are provided in Appendix~\ref{A3}.

\subsubsection{A Simple and Effective Parameterization}

Based on our framework, we show a simple and effective parameter configuration: $a_t = 1-t, b_t = t, \sigma_t^2 = \sigma^2t(1-t)$. Its corresponding sampling ODE can be derived from Eq.~{(\ref{eq.13})} as
\begin{equation}
    \frac{\mathrm{d}\mathbf{x}_t}{\mathrm{d}t} = \frac{1-2t}{2t(1-t)} \mathbf{x}_t - \frac{1}{2t} \mathbf{s} + \frac{1}{2(1-t)} \mathbf{y}.
    \label{eq.18}
\end{equation}
This formulation is known as Brownian bridge (BB) \cite{he2024consistency} or Schr\"odinger bridge-conditional flow matching (SB-CFM) \cite{tong2023improving}, a special case of the SB parameterization listed in Table 1 with $\alpha_t = 1, \rho_t^2 = \sigma^2 t$.

\subsection{Predictive Properties of Flow and Diffusion Bridge Models}

\subsubsection{Predictive Behavior in the Network's Functioning}
Flow matching and diffusion bridge models construct probability paths between data pairs. This contrasts with conventional flow and diffusion models, which typically learn mappings between entire distributions, transforming random samples from a source distribution into samples from a target distribution. Predictive models for SE, by comparison, can be interpreted as implicitly modeling a single-step transition between Dirac distributions centered on the paired data. This perspective aligns with the core objective of the generative models discussed in this paper, highlighting a similarity between these generative approaches and predictive models in terms of their overall processing framework.

\begin{figure}[t]
\centering
\includegraphics[width=0.9\columnwidth]{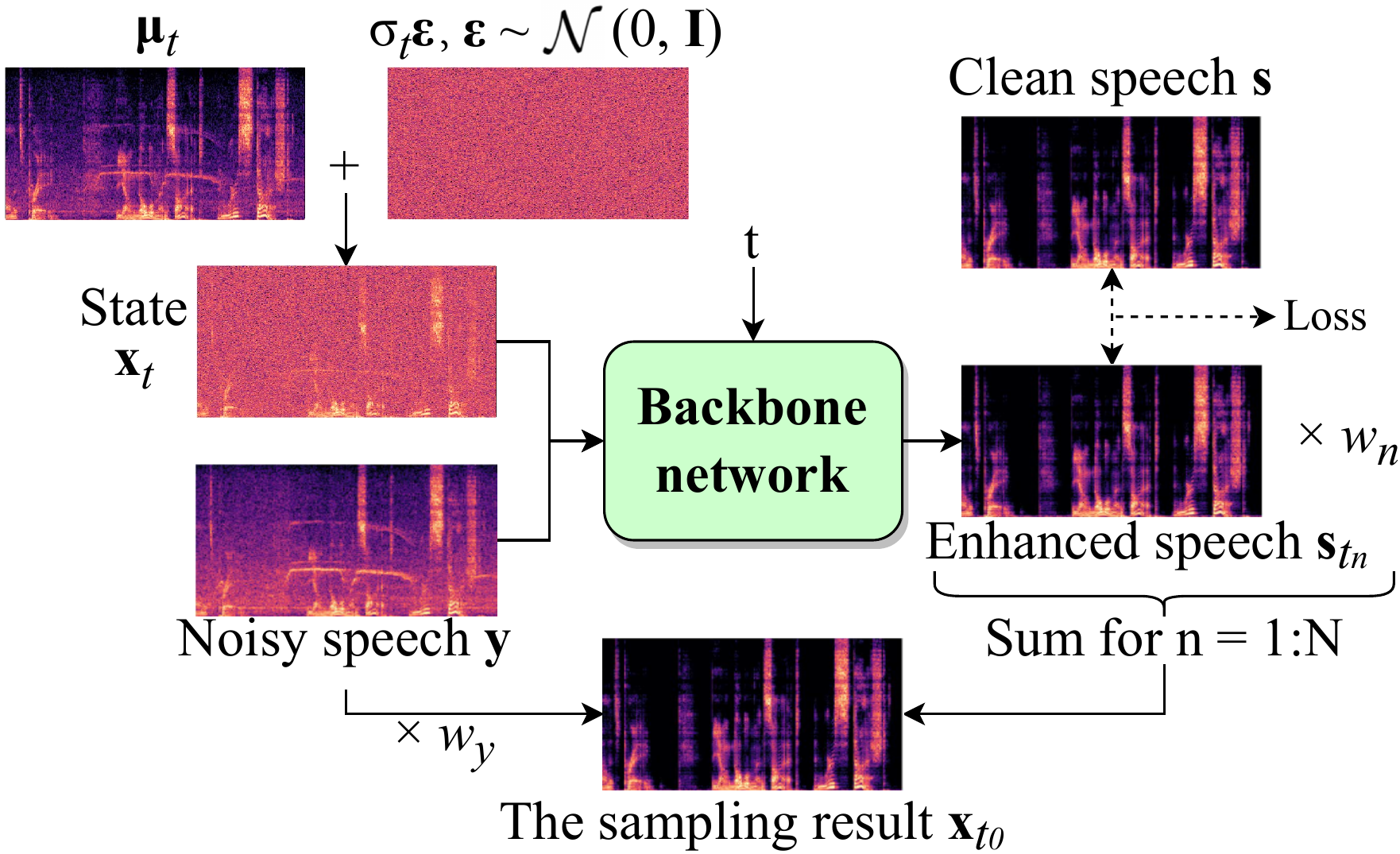}
\caption{Illustration of the backbone network's working mechanism during training and ODE-based sampling (as expressed in Eq.~{(\ref{eq.20})}) under the data prediction strategy.}
\label{fig1}
\end{figure}

Figure~\ref{fig1} illustrates the working mechanism of the backbone network of flow and diffusion bridge models during training and sampling under the data prediction strategy. The network takes as input the state $\mathbf{x}_t$, the noisy signal $\mathbf{y}$, and the time variable $t$, and outputs the enhanced speech. Compared to a standard predictive SE model, two additional inputs, $\mathbf{x}_t$ and $t$, are introduced. The state $\mathbf{x}_t$ follows the Gaussian distribution with mean $\boldsymbol{\mu}_t$ and variance $\sigma_t^2$, where $\boldsymbol{\mu}_t$ is an interpolation between the clean and noisy speech. This makes the mean of $\mathbf{x}_t$ equivalent to a noisy signal with a relatively higher signal-to-noise ratio (SNR). As sampling proceeds, the SNR of $\boldsymbol{\mu}_t$ increases gradually and eventually approaches that of the clean speech. The diffusion time $t$ encodes this SNR progression as well as the level of the variance. Therefore, the backbone network can be viewed as a predictive SE model augmented with auxiliary information. This reveals a strong alignment between the working mechanism of these generative model and conventional predictive models.

\subsubsection{Analysis of Sampling Result Composition}
We analyze the composition of the final sampling result by examining the first-order discretized sampling equation based on the exponential integrator. Specifically, we adopt the first-order discretization of the ODE given in Eq.~{(\ref{eq.17})} and, following the diffusion bridge models, perform sampling in the reverse time direction. For clarity, we rewrite Eq.~{(\ref{eq.17})} as $\mathbf{x}_t = \xi(t, r) \mathbf{x}_r + \eta(t, r) \mathbf{s} + \zeta(t, r) \mathbf{y}$. Substituting the discretized time steps $t = t_n, r = t_{n+1}$, denoting that $\theta(t_n, t_{n+1}) = \theta_n, \theta = \xi, \eta, \zeta$, and replacing the clean speech $\mathbf{s}$ with the network output $\mathbf{s}_{t_{n+1}}$ at each step, the sampling equation can be rewritten as
\begin{equation}
    \mathbf{x}_{t_n} = \xi_n \mathbf{x}_{t_{n+1}} + \eta_n \mathbf{s}_{t_{n+1}} + \zeta_n \mathbf{y}, \mathbf{x}_{t_N} = \mathbf{y}.
    \label{eq.19}
\end{equation}
The final sampling result can then be expressed as 
\begin{equation}
    \mathbf{x}_{t_0} = \sum_{n=1}^N \left( w_n \mathbf{s}_{t_n} \right) + w_y \mathbf{y},
    \label{eq.20}
\end{equation}
where
\begin{equation}
    w_n = \tilde{\xi}_{n-2} \eta_{n-1}, w_y = \sum_{n=1}^{N+1} \tilde{\xi}_{n-2} \zeta_{n-1},
    \label{eq.21}
\end{equation}
with $\tilde{\xi}_{n} = \prod_{k=1}^n \xi_k, n \geq 0, \tilde{\xi}_{-1} = 0$, and $\zeta_N = 1$. It is important to note that the sampling endpoint $t_0$ is typically set to a small positive value (e.g., $10^{-4}$) to avoid numerical singularities. To obtain more specific results, we consider the parameterization of the SB model and apply discretization, obtaining
\begin{equation}
    w_n = \frac{\alpha_0\rho_0\bar{\rho}_0}{\rho_N^2} \left( \frac{\bar{\rho}_{n-1}}{\rho_{n-1}} -  \frac{\bar{\rho}_{n}}{\rho_{n}}\right), w_y = \frac{\alpha_0\rho_0^2}{\alpha_N\rho_N^2}.
    \label{eq.22}
\end{equation}
By substituting the specific parameter values of $\alpha_t$, $\rho_t$, and $\bar{\rho}_t$, the exact values of these weights can be explicitly calculated. Detailed derivations and analyses of the above formulas are provided in Appendix~\ref{A4}.

\begin{figure}[t]
\centering
\includegraphics[width=0.9\columnwidth]{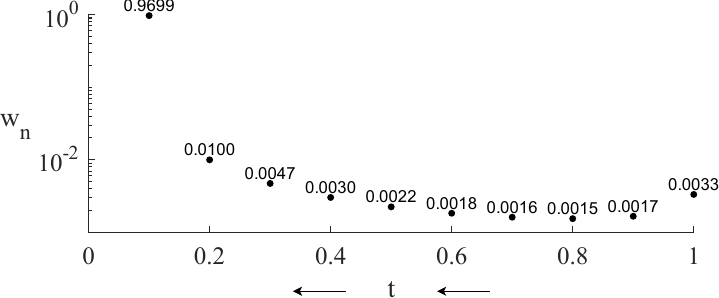}
\caption{Weight distribution of network outputs at each step in ODE-based sampling result (SB-CFM parameterization, and $N = 10$). The arrows indicate that sampling proceeds in the reverse time direction.}
\label{fig2}
\end{figure}

Eq.~(\ref{eq.20}) reveals that the final sampling result is a weighted combination of the network's clean speech estimates at each step and the noisy signal $\mathbf{y}$, with the weights determining their respective contributions. Fig.~\ref{fig1} provides an intuitive illustration of this weighted combination. Using the SB-CFM parameterization (set $\sigma=1$) described above, we perform numerical simulations on the weights defined in Eq.~{(\ref{eq.22})}. Specifically, we set the number of sampling steps $N=10$, obtaining $w_y=10^{-4}$, and the weights $w_n$ at each step are plotted in Fig.~\ref{fig2}. The simulation results indicate that the final output is largely dominated by the network's estimate at the last step, while the contributions from earlier steps and the noisy input $\mathbf{y}$ are negligible. Note that if the network's outputs at each step do not outperform those of traditional predictive models, the SE tasks may not gain a substantial advantage from adopting this generative framework.

Furthermore, it is important to emphasize that one-step sampling is nearly equivalent to a predictive model. Its output relies entirely on a single model call based on data prediction, without leveraging information from intermediate states $\mathbf{x}_t$. In this case, training is only meaningful at $t=1$, while training at other time steps becomes redundant and offers no meaningful contribution to performance.

\subsection{Improved Bridge Model for Speech Enhancement Incorporating Predictive Paradigms}

In the previous section, we analyze the underlying consistency between flow/diffusion bridge models and predictive SE methods. Motivated by this insight, we propose a series of improvements applicable to the flow and diffusion bridge models described by our unified framework. Given the demonstrated advantages of SB parameterization in prior studies, we integrate these enhancements with the SB model to construct an improved bridge model.

\begin{figure}[t]
\centering
\includegraphics[width=1.0\columnwidth]{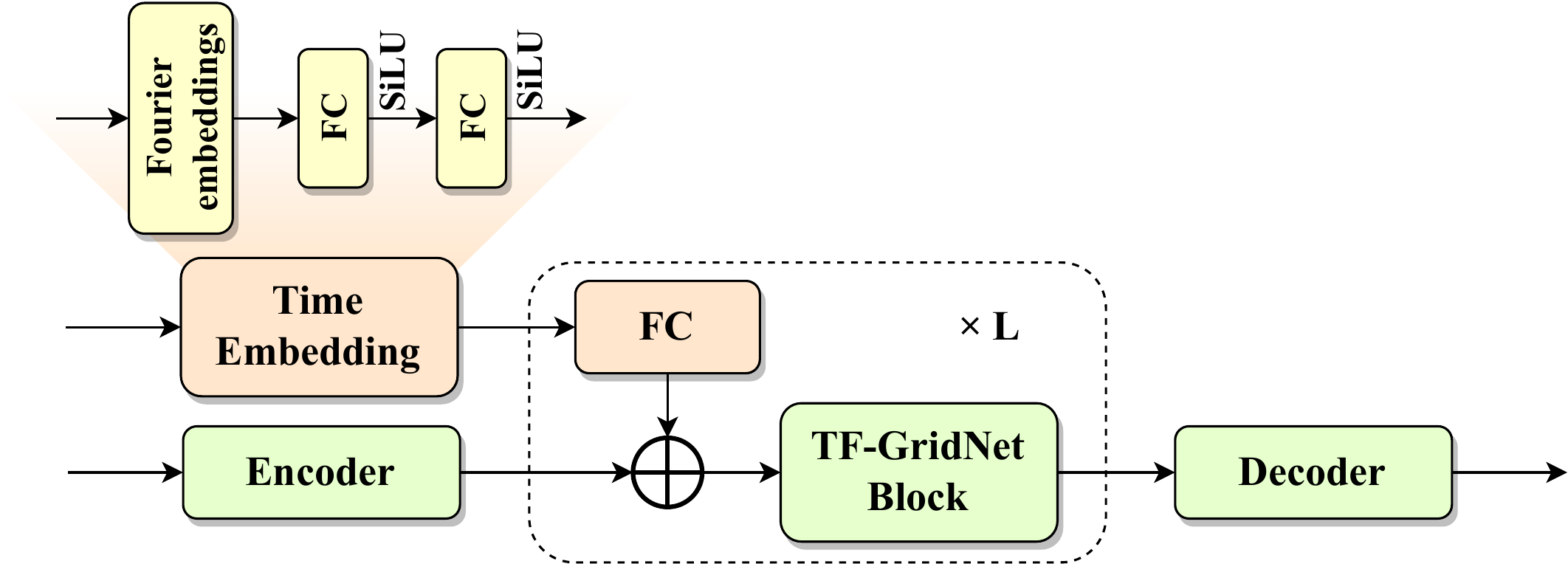}
\caption{Schematic illustration of the time-embedding-assisted TF-GridNet.}
\label{fig3}
\end{figure}

\subsubsection{Improved Backbone Network}
We integrate TF-GridNet \cite{wang2023tf}, a SOTA predictive SE model, as the backbone network in the generative framework, replacing the commonly used U-Net architectures such as Noise Conditional Score Network (NCSN++) \cite{song2020score}. TF-GridNet is highly effective for speech estimation due to its ability to capture correlations between subbands and frames. However, the original TF-GridNet architecture cannot directly accept diffusion time as an input.

To leverage the information in the diffusion time $t$, we introduce a time-embedding mechanism to make TF-GridNet time-dependent. As illustrated in Fig.~\ref{fig1}, the diffusion time is first projected into a high-dimensional vector using a time embedding module, which consists of Fourier embeddings followed by fully connected (FC) layers with sigmoid linear unit (SiLU) activation functions \cite{elfwing2018sigmoid}. The resulting time embedding vector is then incorporated into each TF-GridNet block. Specifically, it is processed by a dedicated FC layer and added to the input features at the start of each TF-GridNet block.

\begin{table*}[t]
\centering
\small
\begin{tabular}{cccc|cc|ccccc}
\Xhline{1pt}
Backbone & Loss & CRP & Schedule & Para. (M) & MACs (G) & SI-SNR & ESTOI & PESQ & DNSMOS & UTMOS \\
\hline\hline
Noisy & - & - & - & - & - & 5.613 & 0.669 & 1.406 & 2.147 & 1.476 \\
\hline
NCSN++ & Original & \scalebox{0.7}{\XSolidBrush} & SBVE & 65.6 & 66 $\times$ 5 & 14.158 & 0.836 & 2.706 & 3.666 & 2.155 \\
NCSN++ & Improved & \scalebox{0.7}{\XSolidBrush} & SBVE & 65.6 & 66 $\times$ 5 & 13.481 & 0.842 & 2.802 & 3.726 & 2.160 \\
TF-GridNet & Improved & \scalebox{0.7}{\XSolidBrush} & SBVE & 2.2 & 38 $\times$ 5 & 16.646 & 0.871 & 3.068 & 3.761 & 2.246 \\
TF-GridNet & Improved & \scalebox{0.7}{\Checkmark} & SBVE & 2.2 & 38 $\times$ 5 & 16.424 & 0.874 & 3.213 & 3.752 & 2.253 \\
\hline
TF-GridNet & Improved & \scalebox{0.7}{\XSolidBrush} & OUVE & 2.2 & 38 $\times$ 60 & 11.302 & 0.778 & 2.129 & 3.385 & 1.874 \\
TF-GridNet & Improved & \scalebox{0.7}{\XSolidBrush} & BBED & 2.2 & 38 $\times$ 60 & 14.429 & 0.843 & 2.800 & 3.691 & 2.133 \\
TF-GridNet & Improved & \scalebox{0.7}{\XSolidBrush} & OT-CFM & 2.2 & 38 $\times$ 5 & 14.866 & 0.851 & 2.834 & 3.385 & 2.168 \\
TF-GridNet & Improved & \scalebox{0.7}{\XSolidBrush} & SB-CFM & 2.2 & 38 $\times$ 5 & 16.177 & 0.867 & 3.102 & 3.742 & 2.216 \\
TF-GridNet & Improved & \scalebox{0.7}{\XSolidBrush} & SBVE & 2.2 & 38 $\times$ 5 & 16.646 & 0.871 & 3.068 & 3.761 & 2.246 \\
\Xhline{1pt}
\end{tabular}
\caption{\label{Table2} Ablation study results on DNS3 test set.}
\end{table*}

\subsubsection{Improved Loss Function}
In previous studies, the data prediction loss for diffusion models is generally defined as a combination of MSE loss on the complex spectrogram, time-domain L1 loss, and PESQ loss. However, these configurations may underemphasize the importance of spectral amplitude and over-optimize PESQ. Therefore, inspired by predictive SE models, we introduce the negative SI-SNR \cite{le2019sdr} loss and the power-compressed specturm loss into the SB-based diffusion model, defined as
\begin{equation}
    \mathcal{L}_{\text{SI-SNR}} \left(\hat{x}, x\right) \! = \! -\log_{10} \! \left(\frac{\left\|x_{t}\right\|^{2}}{\left\|\hat{x} \! - \! x_{t}\right\|^{2}} \! \right) \! , x_{t} \! = \! \frac{\langle\hat{x}, x\rangle x}{\left\|x\right\|^{2}}, 
    \label{eq.23}
\end{equation}
\begin{equation}
    \mathcal{L}_{\text{mag}}\left(\hat{X}, X\right) = \operatorname{MSE}\left(|\hat{X}|^{0.3}, |X|^{0.3}\right),
    \label{eq.24}
\end{equation}
\begin{equation}
    \mathcal{L}_{\text{real/imag}}\left(\hat{X}, X\right) = \operatorname{MSE}\left(\frac{\hat{X}_{\text{r/i}}}{|\hat{X}|^{0.7}}, \frac{X_{\text{r/i}}}{|X|^{0.7}}\right),
    \label{eq.25}
\end{equation}
where $x$ and $\hat{x}$ represent clean and enhanced waveforms, $X$ and $\hat{X}$ are their corresponding spectrograms, the subscripts $\text{r, i}$ represent the real and imaginary parts of the spectrograms, respectively, $\langle \cdot,\cdot \rangle$ denotes the inner product operator, and $\text{MSE}(\cdot,\cdot)$ represents the mean squared error (MSE). The overall loss function for model training is given by
\begin{equation}
    \begin{aligned}
    \mathcal{L} = \lambda_1 \mathcal{L}_{\text{SI-SNR}} + \lambda_2\mathcal{L}_{\text{mag}} + \lambda_3\left(\mathcal{L}_{\text{real}} + \mathcal{L}_{\text{imag}}\right), 
    \end{aligned}
    \label{eq.26}
\end{equation}
where $\lambda_1, \lambda_2, \lambda_3$ are the empirical weights.

\subsubsection{Incorporation of a Predictive Fine-tuning Strategy}
A fine-tuning method called correcting the reverse process (CRP) has been introduced into BBED to mitigate errors accumulated during the sampling process \cite{lay2024single}. CRP fine-tunes the score model by minimizing an MSE loss between the clean speech and the signal generated using the Euler-Maruyama (EuM) first-order sampling method. CRP only updates the model weights during the last model call. This fine-tuning strategy can be generalized to various flow and diffusion bridge models by replacing the EuM method with preferred sampling method, such as the exponential integrator-based approach. Moreover, the original MSE loss used in CRP can be replaced with our improved data prediction loss. It is worth emphasizing that updating weights only at the final step is consistent with our earlier finding that the last sampling step has the greatest influence on the final result and plays a dominant role in the model's overall performance.

\section{Experiments}

\subsection{Experimental setup}

\subsubsection{Datasets and Implementation Details}
We conduct experiments on two datasets. The first dataset is constructed for both denoising and dereverberation tasks, using clean and noise samples from the 3rd Deep Noise Suppression Challenge (DNS3) dataset \cite{reddy2021interspeech}. The second one is the standardized VoiceBank+DEMAND dataset \cite{valentini2016investigating}, which is widely used as a benchmark for SE. All utterances are downsampled from 48 kHz to 16 kHz. Details regarding hyperparameter settings, training configuration, evaluation metrics, and other implementation specifics are provided in Appendix~\ref{B}.

\begin{table*}[t]
\centering
\small
\begin{tabular}{l|cc|ccccc}
\Xhline{1pt}
Model & Para. (M) & MACs (G) & SI-SNR & ESTOI & PESQ & DNSMOS & UTMOS \\ 
\hline\hline
Noisy & - & - & 5.613 & 0.669 & 1.406 & 2.147 & 1.476 \\
\hline
NCSN++ & 59.6 & 66 & 14.146 & 0.842 & 2.673 & 3.747 & 2.182 \\
TF-GridNet & 2.1 & 38 & 16.448 & 0.872 & 3.187 & 3.743 & 2.236 \\
\hline
SGMSE+  & 65.6 & 66 $\times$ 60 & 11.873 & 0.796 & 2.336 & 3.647 & 2.007 \\
StoRM & 65.6 & 66 $+$ 66 $\times$ 60 & 12.463 & 0.805 & 2.297 & 3.625 & 2.060 \\
SBVE & 65.6 & 66 $\times$ 60 & 14.959 & 0.844 & 2.592 & 3.729 & 2.208 \\
\hline
Proposed (NFEs=1) & 2.2 & 38 $\times$ 1 & 16.245 & 0.870 & 3.185 & 3.740 & 2.237 \\
Proposed (NFEs=5) & 2.2 & 38 $\times$ 5 & 16.424 & 0.874 & 3.213 & 3.752 & 2.253 \\
\Xhline{1pt}
\end{tabular}
\caption{\label{Table3} Performance on DNS3 test set.}
\end{table*}

\begin{table}[t]
\centering
\small
\begin{threeparttable}
\begin{tabular}{l|cccc}
\Xhline{1pt}
Model & SI-SNR & ESTOI & PESQ & DNSMOS \\ 
\hline\hline
Noisy & 8.4 & 0.79 & 1.97 & 3.09 \\
\hline
NCSN++ & 18.8 & 0.88 & 3.01 & 3.56 \\
TF-GridNet & 19.5 & 0.88 & 3.17 & 3.57 \\
\hline
SGMSE+\tnote{*} & 17.3 & 0.87 & 2.93 & 3.56 \\
StoRM\tnote{*} & 18.8 & 0.88 & 2.93 & - \\
BBED\tnote{*} & 18.8 & 0.88 & 3.09 & 3.57 \\
SBVE\tnote{*} & 19.4 & 0.88 & 2.91 & 3.59 \\
FlowSE\tnote{*} & 19.0 & 0.88 & 3.12 & 3.58 \\
\hline
Proposed & 19.6 & 0.89 & 3.30 & 3.57 \\
\Xhline{1pt}
\end{tabular}
\begin{tablenotes}
\item[*] Metrics are provided by their original papers.
\end{tablenotes}
\end{threeparttable}
\caption{\label{Table4} Performance on Voicebank+DEMAND test set.}
\end{table}

\subsubsection{Baselines}
We compare the proposed model with several predictive and generative baselines. The predictive baselines include NCSN++ and TF-GridNet, both trained using the proposed loss function. The generative baselines include SGMSE+ (OUVE) \cite{richter2023speech}, StoRM \cite{lemercier2023storm}, BBED \cite{lay2023reducing}, SBVE \cite{jukic2024schrodinger}, and FlowSE \cite{lee2025flowse}. NCSN++ is used as the backbone of SGMSE+, StoRM, and SBVE, following the configuration in \cite{richter2023speech}, resulting in approximately 65.6M parameters. The training and sampling configurations of the baselines follow those of the original papers.

\subsection{Experimental Results}
\subsubsection{Ablation Study Results}
We validate the effectiveness of the proposed modifications on the DNS3 test set. As shown in Table~\ref{Table2}, the ablation study demonstrates that the time-embedding-assisted TF-GridNet along with the improved data prediction loss significantly improves the overall performance of the bridge model, while substantially reducing the number of parameters and computational complexity compared to NCSN++. Additionally, the integration of CRP fine-tuning yields further performance gains without increasing inference cost.

Building on the improved backbone and loss function, we conduct ablation experiments to evaluate several probability path parameterizations, including OUVE, BBED, OT-CFM, SB-CFM, and SBVE, among which only SB-CFM has not been previously applied to SE. Notably, BBED, SB-CFM, and SBVE exhibit zero variance at the starting point of sampling, which corresponds to a Dirac distribution centered on the noisy input. However, due to the complex definitions of $\sigma_t$ in OUVE and BBED,  it is difficult to obtain tractable solutions for the exponential integrator-based samplers. Consequently, we follow the original implementations for OUVE and BBED, employing predictor-corrector (PC) samplers, which require more sampling steps to maintain performance. Experimental results show that SB-CFM and SBVE outperform the alternatives in SE tasks. Based on these findings, we adopt the SBVE schedule as the probability path in our improved bridge model. Overall, the results support the conclusion that Gaussian probability paths with Dirac endpoints, along with exponential integrator-based sampling, provide strong performance guarantees for flow and diffusion bridge models in SE.

\subsubsection{Comparison with the Baseline Models}
The comparison results on the DNS3 test set are presented in Table~\ref{Table3}. Compared with the predictive baselines, the proposed model with one-step sampling outperforms NCSN++ and achieves performance comparable to TF-GridNet, one of the current SOTA predictive models. With additional sampling steps, the proposed model slightly outperforms TF-GridNet across most metrics. This reinforces our earlier conclusion that one-step sampling under this generative framework is essentially equivalent to a predictive model. Furthermore, it significantly surpasses the generative baselines, especially the SOTA SBVE model, in terms of both performance and efficiency, requiring fewer sampling steps, fewer parameters, and lower computational complexity.

Table~\ref{Table4} presents results on the Voicebank+DEMAND test set, with scores for generative baselines taken from their original papers. The proposed model achieves SOTA performance across nearly all metrics, further validating the effectiveness of integrating predictive paradigms into diffusion models. These results also support the view that such generative models inherently exhibit predictive behavior.

\begin{figure}[t]
\centering
\subfigure[]{\includegraphics[height=1.1in]{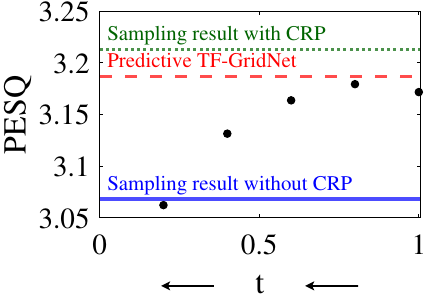}}
\hfil
\subfigure[]{\includegraphics[height=1.1in]{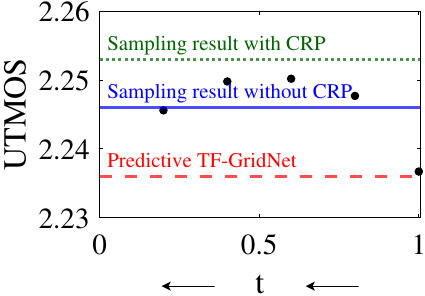}}
\caption{Average PESQ and UTMOS of network outputs at each step during sampling ($N=5$) for the proposed bridge model (without CRP). Dots represent the scores of intermediate network outputs; lines indicate the metrics of the predictive TF-GridNet output and the final sampling results of the proposed bridge model with and without CRP fine-tuning. The arrows indicate that sampling is performed in the reverse time direction.}
\label{fig4}
\end{figure}

\subsubsection{Impact of Predictive Behavior on the Performance of Flow and Diffusion Bridge Models}
Based on our analysis of the inherent equivalence between flow matching/diffusion bridge models and predictive methods, we observe that the quality of the final sampling result is largely determined by the accuracy with which the network estimates the clean speech at each sampling step. Fig.~\ref{fig4} presents the average PESQ and UTMOS of the network outputs at each step ($N=5$) for the proposed bridge model without CRP fine-tuning, along with the scores of the final sampling result. As previously discussed, the network output at the last step ($t = 0.2$) contributes most heavily to the final result, thus leading to nearly identical scores. Fig.~\ref{fig4} also includes the scores of enhanced outputs from the predictive TF-GridNet model, which closely match those of the network output at the first sampling step. This supports our earlier conclusion that at $t = 1$, where the network input consists solely of the noisy signal, the model behaves equivalently to a predictive system.

Furthermore, the scores at all sampling steps are comparable to those of the predictive model, indicating that this generative framework achieves strong denoising and dereverberation performance with each model call. However, this predictive-like behavior suggests an inherent upper bound on performance, that is, it may not significantly outperform its corresponding predictive model for SE tasks. 

Additionally, we observe that during training, the final model call, which dominates the final sampling result, may be slightly under-optimized (lower PESQ than other steps). Fine-tuning this step using CRP compensates for this limitation and further enhances the overall performance of the bridge model.

\section{Conclusion}

In this paper, we present a unified theoretical framework that encompasses widely used generative approaches in SE, including score-based diffusion, Schr\"odinger bridge, and flow matching methods. We demonstrate that these flow and diffusion bridge models, although generative in form, share key mechanisms with predictive SE methods. This insight offers practical guidance for improving such models. Building on this finding, we propose an enhanced bridge model that integrates advanced predictive strategies. Our model achieves significantly better performance and efficiency than existing flow and diffusion baselines. Experimental results further suggest that the inherently predictive behavior of these generative models may impose an upper bound on their performance in denoising and dereverberation tasks.

\section{Acknowledgments}
This work was supported by the National Natural
Science Foundation of China (Grant No. 12274221), the Yangtze River Delta Science and Technology Innovation Community Joint Research Project (Grant No. 2024CSJGG1100), and the AI \& AI for Science Project of Nanjing University.

\bibliography{aaai2026}

\appendix
\numberwithin{equation}{section}
\section{Detailed Derivations and Discussions}

\subsection{Unified Framework for Flow and Diffusion Bridge Models \label{A1}}

We construct a Gaussian probability path between the clean and noisy speech distributions based on the data pair $(\mathbf{s}, \mathbf{y})$:
\begin{equation}
    p_t(\mathbf{x}_t | \mathbf{s}, \mathbf{y}) = \mathcal{N} \left( \mathbf{x}_t; \boldsymbol{\mu}_t(\mathbf{s}, \mathbf{y}) , \sigma_t^2\mathbf{I} \right),
    \label{eq.a1}
\end{equation}
where
\begin{equation}
    \boldsymbol{\mu}_t(\mathbf{x}_t | \mathbf{s}, \mathbf{y}) = a_t \mathbf{s} + b_t \mathbf{y}.
    \label{eq.a2}
\end{equation}
Based on conditional flow matching \cite{lipman2022flow}, the conditional vector field is derived as
\begin{equation}
\begin{aligned}
    &\mathbf{u}_t(\mathbf{x}_t | \mathbf{s}, \mathbf{y}) = \frac{\sigma_t'}{\sigma_t}(\mathbf{x}_t-\boldsymbol{\mu}_t) + \boldsymbol{\mu}_t' \\ & = \frac{\sigma_t'}{\sigma_t} \mathbf{x}_t + \left( a_t' - a_t \frac{\sigma_t'}{\sigma_t} \right) \mathbf{s} + \left( b_t' - b_t \frac{\sigma_t'}{\sigma_t} \right) \mathbf{y}.
    \label{eq.a3}
\end{aligned}
\end{equation}
where the superscript prime indicates the time derivative of the variable. Accordingly, the corresponding ordinary differential equation (ODE) can be expressed as
\begin{equation}
    \frac{\mathrm{d}\mathbf{x}_t}{\mathrm{d}t} = \frac{\sigma_t'}{\sigma_t} \mathbf{x}_t + \left( a_t' - a_t \frac{\sigma_t'}{\sigma_t} \right) \mathbf{s} + \left( b_t' - b_t \frac{\sigma_t'}{\sigma_t} \right) \mathbf{y}.
    \label{eq.a4}
\end{equation}
Using the stochastic differential equation (SDE) extension trick based on the Fokker-Planck equation \cite{holderrieth2025introduction}, the associated forward-backward SDEs are given by
\begin{equation}
    \mathrm{d}\mathbf{x}_t = \left[ \mathbf{u}_t(\mathbf{x}_t | \mathbf{s}, \mathbf{y}) + \frac{1}{2} g_t^2 \nabla_{\mathbf{x}_t} \log p_t(\mathbf{x}_t | \mathbf{s}, \mathbf{y}) \right] \mathrm{d}t + g_t\mathrm{d}\mathbf{w}_t,
    \label{eq.a5}
\end{equation}
\begin{equation}
    \mathrm{d}\mathbf{x}_t = \left[ \mathbf{u}_t(\mathbf{x}_t | \mathbf{s}, \mathbf{y}) - \frac{1}{2} g_t^2 \nabla_{\mathbf{x}_t} \log p_t(\mathbf{x}_t | \mathbf{s}, \mathbf{y}) \right] \mathrm{d}t + g_t\mathrm{d}\bar{\mathbf{w}}_t,
    \label{eq.a6}
\end{equation}
where the score function can be derived as
\begin{equation}
    \nabla_{\mathbf{x}_t} \log p_t(\mathbf{x}_t | \mathbf{s}, \mathbf{y}) = - \frac{\mathbf{x}_t - \boldsymbol{\mu}_t}{\sigma_t^2}.
    \label{eq.a7}
\end{equation}
By substituting Eqs.~{(\ref{eq.a3})(\ref{eq.a7})} into Eqs.~{(\ref{eq.a5})(\ref{eq.a6})}, we obtain
\begin{equation}
    \mathrm{d}\mathbf{x}_t = \left[ \kappa_t^{\scriptscriptstyle +} \mathbf{x}_t + \left( a_t' - a_t \kappa_t^{\scriptscriptstyle +} \right) \mathbf{s} + \left( b_t' - b_t \kappa_t^{\scriptscriptstyle +} \right) \mathbf{y} \right] \mathrm{d}t +  g_t\mathrm{d}\mathbf{w}_t,
    \label{eq.a8}
\end{equation}
\begin{equation}
    \mathrm{d}\mathbf{x}_t = \left[ \kappa_t^{\scriptscriptstyle -} \mathbf{x}_t + \left( a_t' - a_t \kappa_t^{\scriptscriptstyle -} \right) \mathbf{s} + \left( b_t' - b_t \kappa_t^{\scriptscriptstyle -} \right) \mathbf{y} \right] \mathrm{d}t + g_t\mathrm{d}\bar{\mathbf{w}}_t,
    \label{eq.a9}
\end{equation}
with
\begin{equation}
    \kappa_t^{\scriptscriptstyle \pm} = \frac{\sigma_t'}{\sigma_t} \mp \frac{g_t^2}{2\sigma_t^2}.
    \label{eq.a10}
\end{equation}

In this unified framework, we must emphasize that $g_t$ can be chosen arbitrarily, meaning a single probability path may correspond to a family of SDEs with different diffusion coefficients \cite{holderrieth2025introduction}. However, in the derivation of score-based diffusion models and Schr\"odinger bridge models, the forward SDE is typically defined prior to the probability path. Since the forward SDE often uses a fixed drift term, the choice of $g_t$ implicitly determines $\sigma_t$. This means that the predefined SDE used in these models is merely one member of a broader family of equivalent SDEs consistent with the same probability path.

In the following derivation of existing models from our unified framework, we treat $g_t$ defined in these models as an auxiliary parameter to simplify the resulting ODEs and SDEs, particularly to avoid the appearance of the time derivative of $\sigma_t$. We denote this auxiliary parameter as $\tilde{g}_t$ to distinguish it from the true diffusion coefficient, which remains denoted by $g_t$.

\subsection{Universality of the Proposed Framework \label{A2}}

\subsubsection{Score-based Diffusion Models}
For the Ornstein-Uhlenbeck process with variance exploding (OUVE) \cite{lim2023score}, the parameters of its conditional probability path are defined by
\begin{equation}
    a_t = e^{-\gamma t}, b_t = 1 - e^{-\gamma t}, \sigma_t^2 = \frac{c \left( k^{2t} - \mathrm{e}^{-2\gamma t} \right)}{2(\gamma + \log k)}.
    \label{eq.a11}
\end{equation}
We can extend OUVE to more general probability paths with OU-form means, without restricting the specific definition of the variance $\sigma_t^2$.

The auxiliary parameter $\tilde{g}_t$ is given by \cite{sarkka2019applied}
\begin{equation}
    \tilde{g}_t^2 = \left( \sigma_t^2 \right)' + 2 \gamma \sigma_t^2.
    \label{eq.a12}
\end{equation}
In score-based diffusion models, this formula is typically used to derive $\sigma_t$ from a predefined forward SDE. However, within our framework, this relationship is used purely to simplify expressions. Using Eq.~{(\ref{eq.a12})}, we have
\begin{equation}
    \frac{\sigma_t'}{\sigma_t} = \frac{\tilde{g}_t^2}{2\sigma_t^2} - \gamma.
    \label{eq.a13}
\end{equation}
Substituting Eqs.~{(\ref{eq.a11})(\ref{eq.a13})} into Eq.~{(\ref{eq.a4})}, we directly derive the sampling ODE:
\begin{equation}
\begin{aligned}
    \frac{\mathrm{d}\mathbf{x}_t}{\mathrm{d}t} &= \left( \frac{\tilde{g}_t^2}{2\sigma_t^2} - \gamma \right) \mathbf{x}_t - e^{-\gamma t} \frac{\tilde{g}_t^2}{2\sigma_t^2} \mathbf{s} \\&+ \left( \gamma - \frac{\tilde{g}_t^2}{2\sigma_t^2}\left( 1 - e^{-\gamma t} \right) \right) \mathbf{y}.
    \label{eq.a14}
\end{aligned}
\end{equation}
Next, consider the PFODE of the OU-SDE in the original paper \cite{lim2023score}:
\begin{equation}
    \frac{\mathrm{d}\mathbf{x}_t}{\mathrm{d}t} = \mathbf{f}_t(\mathbf{x}_t, \mathbf{y}) - \frac{1}{2} \tilde{g}_t^2 \nabla_{\mathbf{x}_t} \log p_t(\mathbf{x}_t | \mathbf{s}, \mathbf{y}),
    \label{eq.a15}
\end{equation}
where $\mathbf{f}_t(\mathbf{x}_t, \mathbf{y}) = \gamma (\mathbf{y} - \mathbf{x}_t)$. Substituting the drift term and the score function into Eq.~{(\ref{eq.a15})}, we obtain an equivalent ODE form:
\begin{equation}
    \frac{\mathrm{d}\mathbf{x}_t}{\mathrm{d}t} = \gamma (\mathbf{y} - \mathbf{x}_t) + \frac{\tilde{g}_t^2}{2\sigma_t^2}  \left[ \mathbf{x}_t - e^{-\gamma t} \mathbf{s} - (1 - e^{-\gamma t}) \mathbf{y} \right].
    \label{eq.a16}
\end{equation}
This ODE is exactly equivalent to Eq.~{(\ref{eq.a14})}, demonstrating that our unified framework allows direct derivation of the ODE corresponding to an OU process from its probability path.

Subsequently, by applying Eqs.~{(\ref{eq.a8})(\ref{eq.a9})}, we obtain the corresponding forward and backward SDEs:
\begin{equation}
\begin{aligned}
    \mathrm{d}\mathbf{x}_t &= \Bigg[ \left( \frac{\tilde{g}_t^2 - g_t^2}{2\sigma_t^2} - \gamma \right) \mathbf{x}_t - e^{-\gamma t} \frac{\tilde{g}_t^2 - g_t^2}{2\sigma_t^2} \mathbf{s} \\&+ \left( \gamma - \frac{\tilde{g}_t^2 - g_t^2}{2\sigma_t^2}\left( 1 - e^{-\gamma t} \right) \right) \mathbf{y} \Bigg] \mathrm{d}t + g_t\mathrm{d}\mathbf{w}_t,
    \label{eq.a17}
\end{aligned}
\end{equation}
\begin{equation}
\begin{aligned}
    \mathrm{d}\mathbf{x}_t &= \Bigg[ \left( \frac{\tilde{g}_t^2 + g_t^2}{2\sigma_t^2} - \gamma \right) \mathbf{x}_t - e^{-\gamma t} \frac{\tilde{g}_t^2 + g_t^2}{2\sigma_t^2} \mathbf{s} \\&+ \left( \gamma - \frac{\tilde{g}_t^2 + g_t^2}{2\sigma_t^2}\left( 1 - e^{-\gamma t} \right) \right) \mathbf{y} \Bigg] \mathrm{d}t + g_t\mathrm{d}\bar{\mathbf{w}}_t.
    \label{eq.a18}
\end{aligned}
\end{equation}
Since the diffusion coefficient $g_t$ can be chosen arbitrarily, we can set $g_t$ as $\tilde{g}_t$. This simplifies the SDEs to
\begin{equation}
    \mathrm{d}\mathbf{x}_t = \gamma (\mathbf{y} - \mathbf{x}_t) \mathrm{d}t + \tilde{g}_t\mathrm{d}\mathbf{w}_t,
    \label{eq.a19}
\end{equation}
\begin{equation}
\begin{aligned}
    \mathrm{d}\mathbf{x}_t &= \Bigg[ \left( \frac{\tilde{g}_t^2}{\sigma_t^2} - \gamma \right) \mathbf{x}_t - e^{-\gamma t} \frac{\tilde{g}_t^2}{\sigma_t^2} \mathbf{s} \\&+ \left( \gamma - \frac{\tilde{g}_t^2}{\sigma_t^2}\left( 1 - e^{-\gamma t} \right) \right) \mathbf{y} \Bigg] \mathrm{d}t + \tilde{g}_t\mathrm{d}\bar{\mathbf{w}}_t.
    \label{eq.a20}
\end{aligned}
\end{equation}
Clearly, following the same reasoning as used to show the equivalence between Eq.~{(\ref{eq.a14})} and Eq.~{(\ref{eq.a15})}, Eqs.~{(\ref{eq.a19})(\ref{eq.a20})} are fully consistent with the SDEs originally defined in the OUVE model \cite{lim2023score}.

For the Brownian bridge with exponential diffusion coefficient (BBED), the mean of its conditional probability path is defined by \cite{lay2023reducing}
\begin{equation}
    a_t = 1 - t, b_t = t.
    \label{eq.a21}
\end{equation}
As before, we do not specify the form of the variance $\sigma_t^2$, but instead introduce the auxiliary parameter $\tilde{g}_t$, which is given by \cite{sarkka2019applied}
\begin{equation}
    \tilde{g}_t^2 = \left( \sigma_t^2 \right)' + \frac{2\sigma_t^2}{1 - t}.
    \label{eq.a22}
\end{equation}

Following the same procedure as in the previous derivation, we derive the corresponding ODE and SDEs from our unified framework as
\begin{equation}
\begin{aligned}
    \frac{\mathrm{d}\mathbf{x}_t}{\mathrm{d}t} &= \left( \frac{\tilde{g}_t^2}{2\sigma_t^2} - \frac{1}{1-t} \right) \mathbf{x}_t - (1-t) \frac{\tilde{g}_t^2}{2\sigma_t^2} \mathbf{s} \\&+ \left( \frac{1}{1-t} - \frac{\tilde{g}_t^2}{2\sigma_t^2}t \right) \mathbf{y},
    \label{eq.a23}
\end{aligned}
\end{equation}
\begin{equation}
\begin{aligned}
    \mathrm{d}\mathbf{x}_t &= \Bigg[ \left( \frac{\tilde{g}_t^2 - g_t^2}{2\sigma_t^2} - \frac{1}{1-t} \right) \mathbf{x}_t - (1-t) \frac{\tilde{g}_t^2 - g_t^2}{2\sigma_t^2} \mathbf{s} \\&+ \left( \frac{1}{1-t} - \frac{\tilde{g}_t^2 - g_t^2}{2\sigma_t^2} t \right) \mathbf{y} \Bigg] \mathrm{d}t + g_t\mathrm{d}\mathbf{w}_t,
    \label{eq.a24}
\end{aligned}
\end{equation}
\begin{equation}
\begin{aligned}
    \mathrm{d}\mathbf{x}_t &= \Bigg[ \left( \frac{\tilde{g}_t^2 + g_t^2}{2\sigma_t^2} - \frac{1}{1-t} \right) \mathbf{x}_t - (1-t) \frac{\tilde{g}_t^2 + g_t^2}{2\sigma_t^2} \mathbf{s} \\&+ \left( \frac{1}{1-t} - \frac{\tilde{g}_t^2 + g_t^2}{2\sigma_t^2} t \right) \mathbf{y} \Bigg] \mathrm{d}t + g_t\mathrm{d}\bar{\mathbf{w}}_t.
    \label{eq.a25}
\end{aligned}
\end{equation}
By setting $g_t = \tilde{g}_t$, the SDEs are simplified to
\begin{equation}
    \mathrm{d}\mathbf{x}_t = \frac{\mathbf{y} - \mathbf{x}_t}{1 - t} \mathrm{d}t + \tilde{g}_t\mathrm{d}\mathbf{w}_t,
    \label{eq.a26}
\end{equation}
\begin{equation}
\begin{aligned}
    \mathrm{d}\mathbf{x}_t &= \Bigg[ \left( \frac{\tilde{g}_t^2}{\sigma_t^2} - \frac{1}{1-t} \right) \mathbf{x}_t - (1-t) \frac{\tilde{g}_t^2}{\sigma_t^2} \mathbf{s} \\&+ \left( \frac{1}{1-t} - \frac{\tilde{g}_t^2}{\sigma_t^2} t \right) \mathbf{y} \Bigg] \mathrm{d}t + \tilde{g}_t\mathrm{d}\bar{\mathbf{w}}_t.
    \label{eq.a27}
\end{aligned}
\end{equation}
It is straightforward to verify that Eqs.~{(\ref{eq.a23})(\ref{eq.a26})(\ref{eq.a27})} are structurally identical to those presented in the original paper \cite{lay2023reducing}.

\subsubsection{Schr\"odinger Bridge}
The conditional probability path parameters of the Schr\"odinger bridge  are defined as \cite{jukic2024schrodinger}
\begin{equation}
    a_t = \frac{\alpha_t\bar{\rho}_t^2}{\rho_1^2}, b_t = \frac{\bar{\alpha}_t\rho_t^2}{\rho_1^2}, \sigma_t^2 = \frac{\alpha_t^2\bar{\rho}_t^2\rho_t^2}{\rho_1^2},
    \label{eq.a28}
\end{equation}
where $\bar{\alpha}_t = \alpha_t\alpha_1^{-1}$, and $\bar{\rho}_t^2 = \rho_1^2 - \rho_t^2$. This set of formulations do not specify the exact forms of $\alpha_t$ and $\rho_t$, and thus serve as a unified representation for all denoising diffusion bridge models (DDBMs) between paired data \cite{he2024consistency}.

Introducing the auxiliary parameters as
\begin{equation}
    f_t = \frac{\alpha_t'}{\alpha_t}, \tilde{g}_t^2 = \alpha_t^2 \left( \rho_t^2 \right)',
    \label{eq.a29}
\end{equation}
we obtain
\begin{equation}
\begin{aligned}
    &\sigma_t' = f_t \sigma_t + \frac{\tilde{g}_t^2\sigma_t}{2 \alpha_t^2} \left( \frac{1}{\rho_t^2} - \frac{1}{\bar{\rho}_t^2} \right), \\
    &a_t' = a_t f_t - \frac{a_t \tilde{g}_t^2}{\alpha_t^2 \bar{\rho}_t^2}, b_t' = b_t f_t - \frac{b_t \tilde{g}_t^2}{\alpha_t^2 \rho_t^2}.
\end{aligned}
\label{eq.a30}
\end{equation}
Substituting into Eqs.~{(\ref{eq.a4})(\ref{eq.a8})(\ref{eq.a9})}, we derive the following ODE and SDEs:
\begin{equation}
\begin{aligned}
    \frac{\mathrm{d}\mathbf{x}_t}{\mathrm{d}t} &= \left[ f_t + \frac{\tilde{g}_t^2}{2 \alpha_t^2} \left( \frac{1}{\rho_t^2} - \frac{1}{\bar{\rho}_t^2} \right) \right] \mathbf{x}_t \\&- \frac{\tilde{g}_t^2}{2 \alpha_t \rho_t^2} \mathbf{s} + \frac{\bar{\alpha}_t \tilde{g}_t^2}{2 \alpha_t^2 \bar{\rho}_t^2} \mathbf{y},
    \label{eq.a31}
\end{aligned}
\end{equation}
\begin{equation}
\begin{aligned}
    \mathrm{d}\mathbf{x}_t = \Bigg[ \left( f_t + \frac{\bar{\rho}_t^2 (\tilde{g}_t^2 - g_t^2) - \rho_t^2(\tilde{g}_t^2 + g_t^2)}{2 \alpha_t^2 \rho_t^2 \bar{\rho}_t^2} \right) \mathbf{x}_t \\ - \frac{\tilde{g}_t^2 - g_t^2}{2 \alpha_t \rho_t^2} \mathbf{s} + \frac{\bar{\alpha}_t (\tilde{g}_t^2 + g_t^2)}{2 \alpha_t^2 \bar{\rho}_t^2} \mathbf{y} \Bigg] \mathrm{d}t +  g_t\mathrm{d}\mathbf{w}_t,
    \label{eq.a32}
\end{aligned}
\end{equation}
\begin{equation}
\begin{aligned}
    \mathrm{d}\mathbf{x}_t = \Bigg[ \left( f_t + \frac{\bar{\rho}_t^2 (\tilde{g}_t^2 + g_t^2) - \rho_t^2(\tilde{g}_t^2 - g_t^2)}{2 \alpha_t^2 \rho_t^2 \bar{\rho}_t^2} \right) \mathbf{x}_t \\ - \frac{\tilde{g}_t^2 + g_t^2}{2 \alpha_t \rho_t^2} \mathbf{s} + \frac{\bar{\alpha}_t (\tilde{g}_t^2 - g_t^2)}{2 \alpha_t^2 \bar{\rho}_t^2} \mathbf{y} \Bigg] \mathrm{d}t + g_t\mathrm{d}\bar{\mathbf{w}}_t.
    \label{eq.a33}
\end{aligned}
\end{equation}
By taking $g_t = \tilde{g}_t$, the SDEs can be rewritten as
\begin{equation}
    \mathrm{d}\mathbf{x}_t = \Bigg[ \left( f_t - \frac{\tilde{g}_t^2}{\alpha_t^2 \bar{\rho}_t^2} \right) \mathbf{x}_t + \frac{\bar{\alpha}_t \tilde{g}_t^2}{\alpha_t^2 \bar{\rho}_t^2} \mathbf{y} \Bigg] \mathrm{d}t +  \tilde{g}_t\mathrm{d}\mathbf{w}_t,
    \label{eq.a34}
\end{equation}
\begin{equation}
    \mathrm{d}\mathbf{x}_t = \Bigg[ \left( f_t + \frac{ \tilde{g}_t^2}{\alpha_t^2 \rho_t^2} \right) \mathbf{x}_t - \frac{\tilde{g}_t^2}{\alpha_t \rho_t^2} \mathbf{s} \Bigg] \mathrm{d}t + \tilde{g}_t\mathrm{d}\bar{\mathbf{w}}_t.
    \label{eq.a35}
\end{equation}
It is straightforward to confirm that Eqs.~{(\ref{eq.a31})(\ref{eq.a34})(\ref{eq.a35})} are consistent with the original formulation presented in  the SB model \cite{jukic2024schrodinger}.

\subsubsection{Flow matching Methods}
For the flow matching-based models \cite{lee2025flowse, korostik2025modifying}, the derivation of the sampling ODE aligns with our unified framework. The corresponding extension to SDEs can similarly be obtained from Eqs.~{(\ref{eq.a8})(\ref{eq.a9})}. For brevity, we omit the detailed derivation here.

\subsection{Discretized Sampling Equation \label{A3}}

According to the exponential integrator-based discretization method, a low-error discretization of the SDE $\mathrm{d}\mathbf{x}_t = \left( p_t \mathbf{x}_t + m_t \mathbf{s} + n_t \mathbf{y} \right) \mathrm{d}t + g_t\mathrm{d}\bar{\mathbf{w}}_t$, can be expressed as
\begin{equation}
\begin{aligned}
    &\mathbf{x}_t = e ^{\int_r^t p_\tau \mathrm{d}\tau} \mathbf{x}_r + \int_\tau^t e ^{\int_\tau^t p_h \mathrm{d}h} m_\tau \mathbf{s} \mathrm{d}\tau \\&+ \left( \int_\tau^t e ^{\int_\tau^t p_h \mathrm{d}h} n_\tau \mathrm{d}\tau \right) \mathbf{y} + \sqrt{-\int_r^t e ^ {2 \int_\tau^t p_h \mathrm{d}h} g^2(\tau) \mathrm{d}\tau} \boldsymbol{\epsilon},
    \label{eq.a36}
\end{aligned}
\end{equation}
with $\boldsymbol{\epsilon} \sim \mathcal{N}(0, \mathbf{I})$ \cite{chen2023schrodinger}. In data prediction training, $\mathbf{s}$ is estimated by the network at each sampling step and is thus time-dependent. However, assuming that the estimate remains constant over the integration interval allows us to treat $\mathbf{s}$ as a time-invariant quantity and extract it outside the integral in practical sampling.

Due to the tractable form of $p_t$ in the ODE (i.e., $p_t = \frac{\sigma_t'}{\sigma_t}$), the discretized sampling ODE can be simplified as
\begin{equation}
    \mathbf{x}_t = \frac{\sigma_t}{\sigma_r} \mathbf{x}_r + \sigma_t \left( \int_r^t \frac{m_\tau}{\sigma_\tau} \mathrm{d}\tau \right) \mathbf{s} + \sigma_t \left( \int_r^t \frac{n_\tau}{\sigma_\tau} \mathrm{d}\tau \right) \mathbf{y},
    \label{eq.a37}
\end{equation}
where we have leveraged $e ^{\int_r^t p_\tau \mathrm{d}\tau} = \frac{\sigma_t}{\sigma_r}$. This exponential integrator-based discretization helps reduce numerical errors. However, if the remaining two integrals in Eq.~{(\ref{eq.a37})} cannot be solved analytically, it is difficult to employing this discretization for sampling. 

Consider the probability path of optimal transport conditional flow matching (OT-CFM), whose parameters are defined as \cite{korostik2025modifying}
\begin{equation}
    a_t = 1 - t, b_t = t, \sigma_t = (1 - t) \sigma_{\max} + t \sigma_{\min}.
    \label{eq.a38}
\end{equation}
Applying Eq.~{(\ref{eq.a4})}, the corresponding sampling ODE can be expressed by
\begin{equation}
    \frac{\mathrm{d}\mathbf{x}_t}{\mathrm{d}t} = \frac{1}{\sigma_t} \left[ (\sigma_{\min} - \sigma_{\max}) \mathbf{x}_t + \sigma_{\max} \mathbf{s} - \sigma_{\min} \mathbf{y} \right].
    \label{eq.a39}
\end{equation}
Then, the first-order exponential integrator-based discretization of this ODE, derived via Eq.~{(\ref{eq.a37})}, is given by
\begin{equation}
    \mathbf{x}_t = \frac{1}{\sigma_r} \left[ \sigma_t \mathbf{x}_r + \sigma_{\max} (t - r) \mathbf{s} - \sigma_{\min} (t - r) \mathbf{y} \right].
    \label{eq.a40}
\end{equation}
Alternatively, we can directly discretize the ODE using the Euler method:
\begin{equation}
    \mathbf{x}_t = \mathbf{x}_r + \frac{1}{\sigma_r} \left[ (\sigma_{\min} - \sigma_{\max}) \mathbf{x}_r + \sigma_{\max} \mathbf{s} - \sigma_{\min} \mathbf{y} \right] (t - r).
    \label{eq.a41}
\end{equation}
It is straightforward to verify that, for OT-CFM, the exponential integrator-based discretization (Eq.~{(\ref{eq.a40})}) is equivalent to the Euler-based discretization (Eq.~{(\ref{eq.a41})}).

\subsection{Composition of the Final Sampling Result \label{A4}}

To analyze the composition of the final sampling result, we rewrite the first-order discretization of the ODE given in Eq.~{(\ref{eq.a37})} as $\mathbf{x}_t = \xi(t, r) \mathbf{x}_r + \eta(t, r) \mathbf{s} + \zeta(t, r) \mathbf{y}$. By substituting discretized time steps $t = t_n, r = t_{n+1}$, denoting that $\theta(t_n, t_{n+1}) = \theta_n, \theta = \xi, \eta, \zeta$, and replacing the clean speech $\mathbf{s}$ with the network output $\mathbf{s}_{t_{n+1}}$ at each step, the reverse-time sampling equation can be rewritten as
\begin{equation}
    \mathbf{x}_{t_n} = \xi_n \mathbf{x}_{t_{n+1}} + \eta_n \mathbf{s}_{t_{n+1}} + \zeta_n \mathbf{y},
    \label{eq.a42}
\end{equation}
where sampling proceeds from $t_N$ to $t_0$, with $t_0$ typically set to a small positive value (e.g., $10^{-4}$) to avoid numerical singularities. Using this recursive expression, the final sampling result can be written as
\begin{equation}
    \mathbf{x}_{t_0} = \sum_{n=1}^N \left( \tilde{\xi}_{n-2} \eta_{n-1} \mathbf{s}_{t_n} \right) + \left( \sum_{n=1}^{N} \tilde{\xi}_{n-2} \zeta_{n-1} \right) \mathbf{y} + \tilde{\xi}_{N-1} \mathbf{x}_{t_N},
    \label{eq.a43}
\end{equation}
where $\tilde{\xi}_{n} = \prod_{k=1}^n \xi_k, n \geq 0, \tilde{\xi}_{-1} = 0$. For simplicity, we can set $\mathbf{x}_{t_N} = \mathbf{y}$, that is, the stochasticity from the variance at the sampling starting point is neglected. Then we have
\begin{equation}
    \mathbf{x}_{t_0} = \sum_{n=1}^N \left( w_n \mathbf{s}_{t_n} \right) + w_y \mathbf{y},
    \label{eq.a44}
\end{equation}
where
\begin{equation}
    w_n = \tilde{\xi}_{n-2} \eta_{n-1}, w_y = \sum_{n=1}^{N+1} \tilde{\xi}_{n-2} \zeta_{n-1},
    \label{eq.a45}
\end{equation}
with $\zeta_N = 1$. This reveals that the final sampling result is a weighted combination of the network's clean speech estimates across all steps and the noisy signal $\mathbf{y}$, with the weights determining their respective contributions. 

For the SB parameterization, the discretized ODE based on exponential integrators is given by
\begin{equation}
\begin{aligned}
    \mathbf{x}_{t_n} =& \frac{\alpha_n \rho_n \bar{\rho}_n}{\alpha_{n+1} \rho_{n+1} \bar{\rho}_{n+1}} \mathbf{x}_{t_{n+1}} + \frac{\alpha_n}{\rho_N^2} \left(\bar{\rho}_n^2 - \frac{\bar{\rho}_{n+1} \rho_n \bar{\rho}_n}{\rho_{n+1}}\right) \mathbf{s}_{t_{n+1}} \\&+ \frac{\alpha_n}{\rho_N^2\alpha_N}\left(\rho_n^2 - \frac{\rho_{n+1} \rho_n \bar{\rho}_n}{\bar{\rho}_{n+1}}\right) \mathbf{y}
    \label{eq.a46}
\end{aligned}
\end{equation}
Comparing Eq.~{(\ref{eq.a42})} and Eq.~{(\ref{eq.a46})}, and substituting the coefficients into Eq.~{(\ref{eq.a45})}, we obtain
\begin{equation}
    w_n = \frac{\alpha_0\rho_0\bar{\rho}_0}{\rho_N^2} \left( \frac{\bar{\rho}_{n-1}}{\rho_{n-1}} -  \frac{\bar{\rho}_{n}}{\rho_{n}}\right), w_y = \frac{\alpha_0\rho_0^2}{\alpha_N\rho_N^2},
    \label{eq.a47}
\end{equation}
with $n = 1, \cdots, N$. In many common SB configurations, $\alpha_t \equiv 1$, $\rho_0 \approx 0$, and $\rho_N$ is significantly larger than $\rho_0$. Therefore, $w_y \approx 0$, which indicates that the noise signal $\mathbf{y}$ contributes minimally to the final output. Additionally, $\sum_{n=1}^N w_n = \alpha_0 \left( 1 - \frac{\rho_0^2}{\rho_N^2} \right) \approx 1$, implying that if the network estimates are accurate, the final output maintains the amplitude of the clean speech signal

\section{Details of the Experimental Setup \label{B}}

\subsection{Datasets}

The two datasets used in the experiments include the DNS3 dataset and the VoiceBank+DEMAND dataset. The DNS3 dataset is constructed using clean and noise samples from the 3rd Deep Noise Suppression Challenge (DNS3) database \cite{reddy2021interspeech}. Clean speech signals are convolved with randomly selected room impulse responses (RIRs) and then mixed with randomly chosen noise clips at SNRs ranging from -5 dB to 15 dB. The training target is generated by preserving only the first 100 ms of reverberation. A total of 72,000 pairs of 10-second noisy-clean utterances are created for training, while 1,000 pairs are generated for validation and testing, respectively. The VoiceBank+DEMAND dataset \cite{valentini2016investigating} is widely used as a benchmark for SE. A total of 1,000 samples are randomly selected from its training set to serve as the validation set. All utterances are downsampled from 48 kHz to 16 kHz.

\subsection{Implementation Details}

The short-time Fourier transform (STFT) is computed with a segment length of 32 ms, an overlap of 50\%, a fast Fourier transform (FFT) length of 512, and a square-root Hann window for analysis and synthesis, with a sample rate of 16 kHz. We adopt the amplitude-compressed STFT representation as the network input \cite{richter2023speech}. For OUVE, BBED, SBVE, and OT-CFM schedules, we adopt the same hyperparameter settings as reported in their respective original papers. For SB-CFM, we set $\sigma = 1$.

For time-embedding-assisted TF-GridNet, we employ 5 stacked blocks with 32 embedding dimensions, 100 hidden units in LSTM, and unfold layers configured with a kernel size of 4 and a stride of 1. The time embedding module maps the diffusion time to a 64-dimensional vector, which is further projected to 128 dimensions by the FC layers. In each TF-GridNet block, the time embedding features are downsampled via the dedicated FC layer to match the dimensions of the hidden features. For the loss function, we set the weights as follows: $\lambda_1 = 0.01$, $\lambda_2 = 0.7$, and $\lambda_3 = 0.3$. For the CRP method integrated into our model, we adopt similar parameter settings to those in \cite{lay2024single}, except for the proposed modifications to the sampler and loss function. For our proposed model, we adopt the ODE sampler with 5 sampling steps unless otherwise specified.

For model training, we employ the same ADAM optimizer and exponential moving average (EMA) configurations as in \citeauthor{richter2023speech} \citeyear{richter2023speech}. During the first training stage, a linear-warmup cosine-annealing scheduler is used. Specifically, the learning rate linearly increases from $5 \times 10^{-6}$ to $5 \times 10^{-4}$ over the first 20,000 steps and then decays following a cosine schedule until 200,000 steps. For the CRP fine-tuning phase, the learning rate is initialized at $1 \times 10^{-4}$ and decays with a factor of 0.99995 at each step. All models are trained on four NVIDIA RTX 4090 GPUs with a batch size of 16.

\subsection{Evaluation Metrics}

We employ a set of commonly used speech objective metrics for evaluation, including SI-SNR \cite{le2019sdr}, wide-band PESQ \cite{rix2001perceptual}, extended short-time objective intelligibility (ESTOI) \cite{jensen2016algorithm}, DNSMOS P.808 \cite{reddy2021dnsmos}, and UTMOS \cite{saeki2022utmos}. Higher values indicate better performance for all metrics.

\end{document}